\documentclass[aps,prd,twocolumn,nofootinbib,superscriptaddress]{revtex4}
\usepackage{latexsym}
\usepackage{hyperref}
\usepackage{amssymb}
\usepackage{epsfig,amsmath,graphics}
\usepackage{pstricks}
\usepackage{color}
\usepackage{dcolumn}
\usepackage{ulem}
\usepackage{xcolor}
\usepackage{placeins}

\definecolor{linkcolor}{rgb}{0.0, 0.28, 0.67}
\hypersetup{colorlinks=true,
linkcolor=black,
citecolor=linkcolor,
urlcolor=linkcolor,
linktocpage=true
}

\newcommand{\TeV}{\text{TeV}}

\newcommand{\eV}{\text{eV}}

\newcommand{\keV}{\text{keV}}

\newcommand{\order}[1]{\mathcal{O}{(#1)}}
\newcommand{\D}{\cal{D}}
\newcommand{\ksm}{\text{km} / \text{s} / \text{Mpc}}
\newcommand{\QDMAP}{Q_\text{DMAP}}

\newcommand{\Eq}[1]{Eq.~(\ref{eq:#1})}
\newcommand{\Fig}[1]{Fig.~\ref{fig:#1}}
\newcommand{\Sec}[1]{Sec.~\ref{sec:#1}}
\newcommand{\App}[1]{Appendix~\ref{app:#1}}
\newcommand{\Tab}[1]{Table~\ref{tb:#1}}

\newcommand{\nir}{N_\text{IR}}
\newcommand{\nuv}{N_\text{UV}}
\newcommand{\nfs}{N_\text{fs}}

\usepackage{comment}
\newcommand{\be}{\begin{equation}}
\newcommand{\ee}{\end{equation}}

\def\bea{\begin{eqnarray}}
\def\eea{\end{eqnarray}}
\def\ltap{\ \raise.3ex\hbox{$<$\kern-.75em\lower1ex\hbox{$\sim$}}\ }
\def\gtap{\ \raise.3ex\hbox{$>$\kern-.75em\lower1ex\hbox{$\sim$}}\ }
\def\lsim{\ \raise.3ex\hbox{$<$\kern-.75em\lower1ex\hbox{$\sim$}}\ }
\def\gsim{\ \raise.3ex\hbox{$>$\kern-.75em\lower1ex\hbox{$\sim$}}\ }

\usepackage{slashed}

\usepackage{color}

\newcommand{\zt}{z_t}

\newcommand{\ignore}[1]{}
\newcommand{\beq}{\begin{equation}}
\newcommand{\eeq}{\end{equation}}
\newcommand{\bear}{\begin{eqnarray}}
\newcommand{\eear}{\end{eqnarray}}

\def\MeV{\,{\rm MeV}}

\def\ev{\,{\rm eV}}
\def\lcdm{$\Lambda{\rm CDM}$}

\newcommand{\Neff}{N_{\rm eff}}

\newcommand{\LCDM}{\Lambda{\rm CDM}}

\begin{document}

\title{A Step in Understanding the Hubble Tension}

\author{Daniel Aloni}
\affiliation{Physics Department, Boston University, Boston, MA 02215, USA}
\author{Asher Berlin}
\affiliation{Center for Cosmology and Particle Physics, Department of Physics, New York University, New York, NY 10003, USA}
\author{Melissa Joseph}
\affiliation{Physics Department, Boston University, Boston, MA 02215, USA}
\author{Martin Schmaltz}
\affiliation{Physics Department, Boston University, Boston, MA 02215, USA}
\author{Neal Weiner}
\affiliation{Center for Cosmology and Particle Physics, Department of Physics, New York University, New York, NY 10003, USA}
\begin{abstract}

As cosmological data have improved, tensions have arisen. One such tension is the difference between the locally measured Hubble constant $H_0$  and the value inferred from the cosmic microwave background (CMB). Interacting radiation has been suggested as a solution, 
but studies show that conventional models are precluded by high-$\ell$ CMB polarization data. It seems at least plausible that a solution may be provided by related models that distinguish between high- and low-$\ell$ multipoles. When interactions of strongly-coupled radiation are mediated by a force-carrier that becomes non-relativistic, the dark radiation undergoes a ``step'' in which its relative energy density increases as the mediator deposits its entropy into the lighter species. If this transition occurs while CMB-observable modes are inside the horizon, high- and low-$\ell$ peaks are impacted differently, corresponding to modes that enter the horizon before or after the step.  These dynamics are naturally packaged into the simplest supersymmetric theory, the Wess-Zumino model, with the mass of the scalar mediator near the $\eV$-scale. We investigate the cosmological signatures of such ``Wess-Zumino Dark Radiation" (WZDR) and find that it provides an improved fit to the CMB alone, favoring larger values of $H_0$. If supernovae  measurements from the SH0ES collaboration are also included in the analysis, the inferred value of $H_0$ is yet larger, but the preference for dark radiation and the location of the transition is left nearly unchanged. Utilizing a standardized set of measures, we compare to other models and find that WZDR is among the most successful at addressing the $H_0$ tension and the best of those with a Lagrangian formulation.

\end{abstract}

\maketitle

\section{Introduction}
The first results from WMAP harkened the arrival of the era of precision cosmology. The many telescopes - terrestrial and orbital - since then have allowed us to test the detailed nature of the evolution of the Universe. While this has given us incredibly precise measurements of the parameters of \lcdm, it has also allowed us to study whether small deviations from standard cosmology might appear.

As the data have come in, some tensions have begun to emerge. The most notable of these is the disagreement between the value of $H_0$ inferred from Planck observations of the CMB~\cite{Planck:2018vyg} and the value extracted from late-universe supernovae data by the SH0ES collaboration~\cite{Riess:2020fzl}. Indeed, it has been argued that the early-universe inferred value shows a clear distinction from a broad set of late-universe measurements~\cite{Riess:2020fzl,Freedman:2019jwv,Freedman:2021ahq,Yuan:2019npk,Soltis:2020gpl,Wong:2019kwg,Pesce:2020xfe,LIGOScientific:2019zcs}.\footnote{There are a variety of measurements of $H_0$, including alternative calibrations of supernovae. To directly compare to Ref.~\cite{Schoneberg:2021qvd}, we focus on the $H_0$ value inferred from SH0ES~\cite{Riess:2020fzl}.} At the same time, there is a well-known, smaller tension with \lcdm\ within the Planck CMB data itself; compared to the low-$\ell$ multipoles, the high-$\ell$ multipoles in the TT power spectrum prefer smaller values of $H_0$ (and larger values of the matter density $\Omega_m h^2$)~\cite{Addison:2015wyg,Planck:2018vyg}. 

There is no clear solution to these tensions. It has been argued that the discrepancy in $H_0$ is due to a deviation in the sound horizon of \lcdm~\cite{Bernal:2016gxb,Aylor:2018drw}. As a consequence, many proposed solutions involve the addition of new components to the cosmological energy density before matter-radiation equality, which has the effect of decreasing the sound horizon at the time of recombination, thus increasing the Hubble rate inferred from the CMB (see, e.g., Refs.~\cite{Brust:2017nmv,Blinov:2020hmc,RoyChoudhury:2020dmd,Brinckmann:2020bcn,Kreisch:2019yzn,Escudero:2019gvw,EscuderoAbenza:2020egd,Escudero:2021rfi,Karwal:2016vyq,Poulin:2018cxd,Lin:2019qug,Smith:2019ihp,Bansal:2021dfh,Cyr-Racine:2021alc,Niedermann:2020dwg} for a representative set).  Additional radiation is one such possibility~\cite{Bernal:2016gxb, Blinov:2020hmc, RoyChoudhury:2020dmd,Brinckmann:2020bcn,Kreisch:2019yzn}, although at the cost of suppressing power at high multipoles of the CMB due to enhanced Silk damping~\cite{Hou:2011ec}. The situation is worsened if this radiation is free-streaming, since in this case the acoustic peaks in the CMB angular power spectrum are additionally suppressed in their amplitude as well as shifted in their location to smaller multipoles   by the so-called ``drag effect"~\cite{Bashinsky:2003tk,Baumann:2015rya,Follin:2015hya}. The drag effect is significantly suppressed if the radiation is instead strongly-coupled to itself, which allows for a more sizable shift to the sound horizon in models of interacting radiation. However, high-$\ell$ polarization measurements of the CMB disfavor models involving enough additional radiation to substantially increase $H_0$~\cite{Bernal:2016gxb, Blinov:2020hmc, RoyChoudhury:2020dmd, Brinckmann:2020bcn}, leaving an open question of whether such basic frameworks can help explain the data.

Motivated by the low- versus high-$\ell$ determinations of \lcdm\ parameters and the incompatibility of substantial amounts of interacting radiation with high-$\ell$ polarization data, we examine in this paper an $\ell$-dependent solution to the Hubble tension. Specifically, we seek to understand whether cosmological and astrophysical data favor strongly-coupled radiation with a mass-threshold near the eV-scale, which increases the relative density in such radiation at late times. Our main result is that including such ``stepped'' dark radiation does significantly improve the combined fit to CMB, BAO, and SH0ES data while not degrading the fit to the CMB and BAO. We find that when fit to a dataset including CMB and BAO (but no direct late-universe measurements of $H_0$), our model predicts $H_0 = [67.6,70.9] \ \ksm$ at 90\% confidence (in good agreement with Ref.~\cite{Freedman:2019jwv}) and reduces the $4.5 \sigma$ tension between \lcdm\ and SH0ES to $2.7 \sigma$. To further quantify the ability of the model to resolve the tension, we apply the ``$H_0$ Olympics'' measures defined in Ref.~\cite{Schoneberg:2021qvd} and find that it is the only concrete particle physics model (with a Lagrangian) that passes the established rubric in all three Olympic criteria (thus qualifying for a ``gold medal").

In \Sec{model}, we describe the overall setup and outline the basic physics of a stepped dark sector, finding that it naturally fits into the simplest supersymmetric model, the Wess-Zumino model. In \Sec{step}, we describe the effects of a mass-threshold on the CMB. In \Sec{data}, we perform various fits to the data, both including and not including late-time determinations of $H_0$.  In \Sec{4param}, we generalize this model to one with a larger set of parameters that allows for a broader consideration of other related theories. Finally, in  \Sec{conclusion} we discuss how this model is favored by the data and conclude. We also provide two appendices that include details of our implementation of the mass-threshold and a full set of posterior densities and best-fit cosmological parameters as determined by our MCMC analysis.

\section{A Model of Stepped Dark Radiation}
\label{sec:model}

Constructing a model of interacting radiation a priori is not difficult. A Weyl fermion $\psi$ and complex scalar $\phi$ that interact through a Yukawa coupling, $\lambda \,  \psi^2 \phi$, provide a compact and economical model of interacting radiation. Scalars are naturally massive with masses near the cutoff of the effective theory, and thus integrating out $\phi$ yields a four-fermion operator, $\lambda^2 \psi^4 / m_\phi^2$. Requiring these interactions to be faster than a Hubble time at recombination bounds the mass of the scalar from above, $m_\phi/\lambda \lesssim \MeV$ (assuming that the dark radiation temperature $T_d$ is comparable to the neutrino temperature). The presence of a light scalar (which is not a Goldstone boson), while efficient practically, is conceptually incomplete and calls for additional physics to stabilize its mass.

Remarkably, the above picture follows from the simplest possible supersymmetric model, the Wess-Zumino model~\cite{Wess:1974tw}, which has the benefit of automatically controlling the scalar mass. The fermion and scalar components of a single superfield $X$ with a superpotential coupling $W= (\lambda / 3) \, X^3$ yields the Lagrangian
\be
{\cal L}_\text{WZ} =  \lambda \, \phi \, \psi^2 + \lambda^2 \, (\phi^* \phi)^2
~.
\ee
We refer to this specific model as ``Wess-Zumino Dark Radiation'' or WZDR. 

Since supersymmetry cannot be exact in Nature, we expect an additional scalar mass $m_\phi^2 \, \phi^* \phi$, which may be generated from interactions with, e.g., the Standard Model. Even without couplings to the Standard Model, gravity will naturally generate scalar masses of order $m_\phi \sim (M_\text{SUSY})^2/M_\text{Pl}$, where $M_\text{SUSY}$ is the fundamental supersymmetry breaking scale and $M_\text{Pl}$ is the Planck mass. For example, in low-scale gauge mediation~\cite{Dine:1993yw,Dine:1995ag} $M_\text{SUSY}\sim100 \ \TeV$ and intriguingly the scalar mass is comparable to the temperature of the photon-baryon plasma at recombination,  $m_\phi \sim 1 \ \ev$.\footnote{If $X$ is sequestered from SUSY breaking, anomaly mediation~\cite{Randall:1998uk,Giudice:1998xp} generates a scalar mass $m_\phi \sim \lambda^2/(16 \pi^2)\, M_\text{SUSY}^2/M_\text{Pl}$.} Current and future measurements of the CMB are and will be sensitive to new relativistic species during the epoch when the temperature of the baryon-photon plasma was $T_\gamma \sim (0.5 - 50) \ \eV$. This happy coincidence, that supersymmetric models of interacting radiation often predict a mass-threshold that occurs within a detectable epoch, makes them distinguishable from conventional models of strongly-coupled radiation.

\begin{figure}[t]
   \centering
   \includegraphics[width=0.5\textwidth]{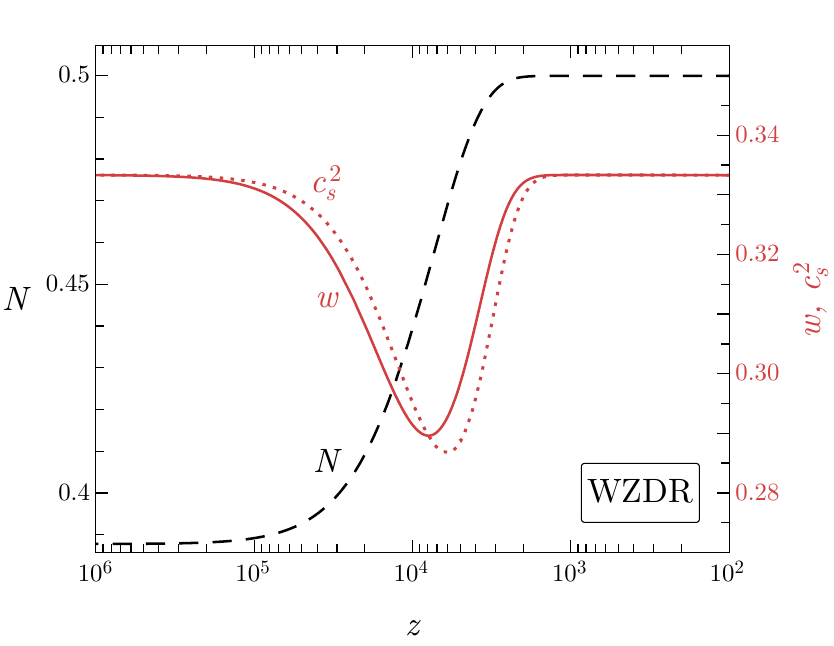}
   \caption{A representative example of the redshift dependence of the effective number of additional neutrinos $N$ (dashed black, left-axis), the equation of state $w$ (solid red, right-axis), and the speed of sound $c_s^2$ (dotted red, right-axis) for a strongly-coupled WZDR fluid that transitions at a redshift of $z_t = 2 \times 10^4$.}
   \label{fig:csofa}
\end{figure}

This then presents a physical picture as follows: some process (examples of which we will discuss below) produces an amount of interacting radiation consisting of $\psi$ and $\phi$ particles with an energy density equivalent to $\nuv$ additional neutrino species after Big Bang nucleosynthesis (BBN). When the temperature of this sector drops below $m_\phi$, the scalars deposit their entropy into the lighter $\psi$ species as decays and annihilations between $\phi$ and $\psi$ maintain chemical and kinetic equilibrium. 
For temperatures not too far below $m_\phi$, the fluid is a mixture of both massive and massless particles such that its energy density redshifts more slowly than that of the relativistic neutrinos. As a result, over a period of time spanning approximately a decade in redshift, the relative energy density of the fluid, as quantified by the effective number of additional neutrinos species $N(z)$, increases to a value $\nir > \nuv$.
The size of this increase follows from conservation of comoving entropy, which gives
\be
\label{eq:step}
\frac{\nir}{\nuv }
= \left(\frac{g_*^\phi +g_*^\psi}{g_*^\psi}\right)^{1/3} =\left(\frac{15}{7}\right)^{1/3}\simeq 1.29
~,
\ee
where $g_*^{\phi, \psi}$ is the effective number of relativistic degrees of freedom in $\phi$, $\psi$, respectively.\footnote{\Eq{step} quickly follows from the fact that conservation of entropy implies that below the mass-threshold, the fluid temperature (relative to the neutrino bath) increases by a factor of $(1+g_*^\phi/g_*^\psi)^{1/3}$, whereas the number of relativistic degrees of freedom decreases by $g_*^\psi/(g_*^\psi+ g_*^\phi)$.} We refer to this general picture in which the radiation density increases as a ``stepped'' model, because the effective amount of interacting radiation relative to neutrinos at the same epoch, $N(z)$, increases from $\nuv$ at large redshift $z$ to $\nir$ at low $z$ as the fluid transitions through the mass-threshold. This model is parametrized by two quantities: $\nir$, the amount of interacting radiation at late times, and $\zt$, the redshift at which the transition occurs (see \App{background} for the precise definition of $z_t$). The amount of interacting radiation at early times $\nuv$ is determined by  \Eq{step}.
Note that at temperatures well below $m_\phi$, when the thermal abundance of $\phi$ is exponentially suppressed, the remaining fermions $\psi$ continue to interact via the four-Fermi interaction mediated by virtual $\phi$'s. Due to the small $\phi$ mass, this interaction rate naturally remains much larger than the Hubble rate well into the era of matter domination, implying that the remaining $\psi$ particles continue to behave as a perfect fluid. 

The transition from $\nuv \to \nir$ is not instantaneous, but rather occurs over a period of time. As the temperature $T_d$ drops below the mass of $\phi$, its energy density becomes exponentially suppressed. Entropy conservation dictates how $T_d$ and the energy density of the dark sector evolve as a function of the scale factor. We note that simply mapping between two relativistic fluids before and after the transition is not enough to understand the detailed evolution of its perturbations. During the transition the fluid is a mixture of massive and massless particles; consequently, its equation of state $w$ lies somewhere between that of a massive ($w=0$) and massless ($w=1/3$) particle. Likewise, the deviation of the sound speed $c_s$ away from that of a relativistic fluid value ($c_s^2 = 1/3$) during the transition leaves a non-negligible imprint at the level of cosmological perturbations. Since the transition occurs while $\phi$ and $\psi$ are chemically coupled, we use entropy conservation to numerically determine the dark sector temperature as a function of redshift, $T_d(z)$, which can then be used as an input to calculate the redshift evolution of $N$, $w$, and $c_s^2$, as shown in  \Fig{csofa}. Further details are provided in \App{background}. 

\section{The effects of a Step on the CMB}
\label{sec:step}

\begin{figure}[t]
\centering
\includegraphics[width=0.5\textwidth]{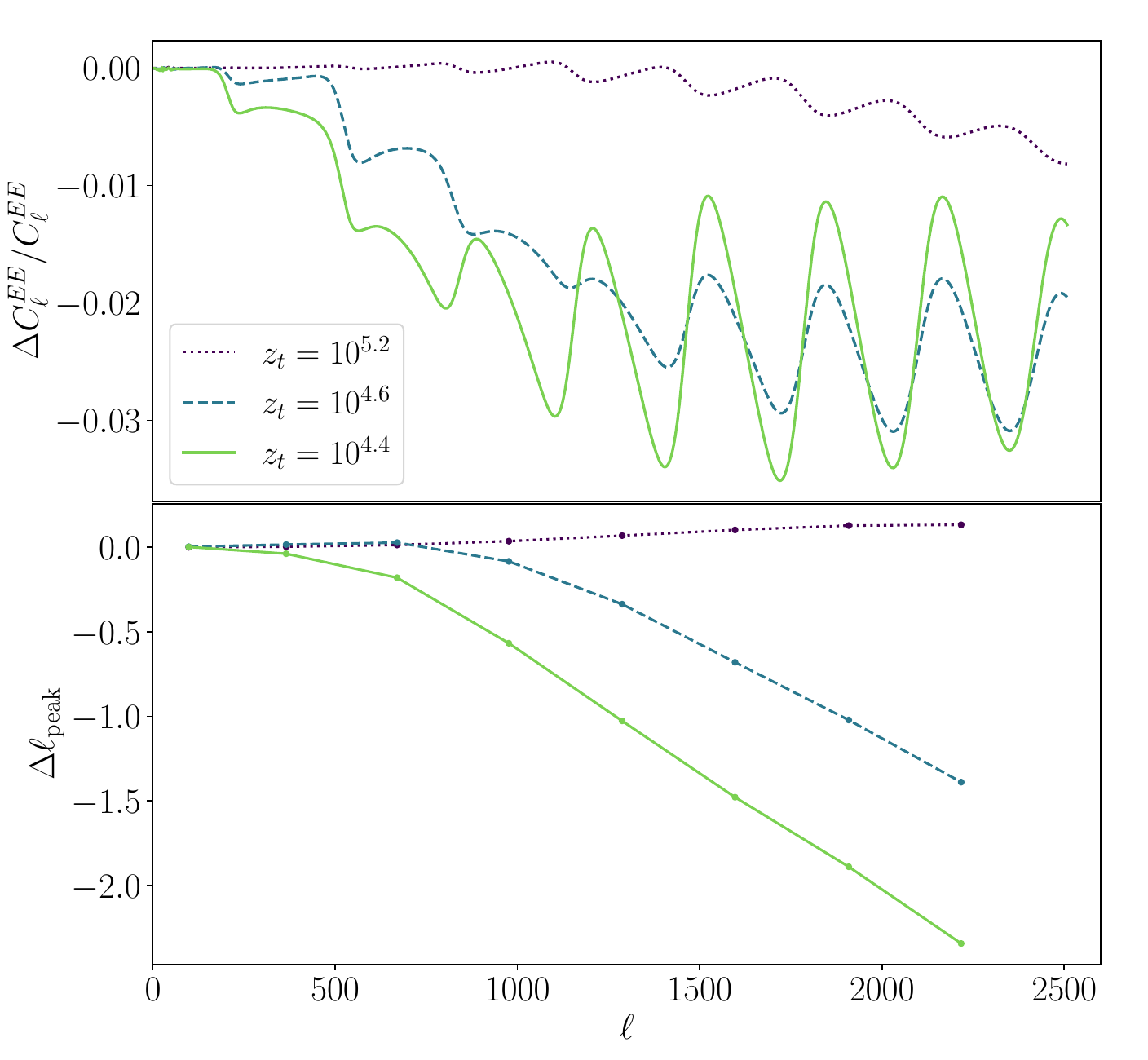} 
\caption{({\it Top}) The relative change to the EE spectrum $\Delta C_\ell^{EE} / C_\ell^{EE}$ between the WZDR model and a reference SIDR model for a variety of step locations $z_t$. ({\it Bottom}) The shift to the position of the peaks $\Delta \ell_\text{peak}$ in the EE power spectrum as a function of $\ell$ for the same choices of $z_t$. In each case, we plotted the lensed spectrum for the $\LCDM$ parameters $H_0=71.35\,$km/s/Mpc, $\omega_{\rm b}=0.02272$, $\omega_{\rm CDM}=0.1288$, $\tau_{\rm reio}=0.0586$, $N_{\nu}=3.044$, $\sum m_\nu=0.06\,$eV, $Y_p=0.2455$, and the energy density in the new fluid is $\nir = 0.6$. The effects are qualitatively similar for the TT spectrum.}
\label{fig:clwithstep}
\end{figure}

Before delving into a global analysis of the impact of WZDR on the data, let us briefly summarize the imprints of a stepped fluid on the CMB (for a discussion of stepped fluids with earlier transitions and their effects on BBN, see Ref.~\cite{Berlin:2019pbq}). Perhaps the simplest comparison to make is between a stepped and unstepped interacting fluid in which late-universe parameters are matched. This is illustrated in \Fig{clwithstep}, which shows the relative change in the CMB angular power spectrum $\Delta C_\ell / C_\ell$ (top-panel) and the change in the CMB peak position $\Delta \ell_\text{peak}$ (bottom-panel) between a model in which the energy density of interacting fluid transitions as $\nuv \to \nir$ at a redshift $\zt$, and a reference model in which the energy density of interacting fluid is always $\nir$. In this comparison, we match all other cosmological input parameters, such as $\Omega_\Lambda$, $\Omega_m$, and $Y_p$. We refer to the reference model as ``self-interacting dark radiation" or SIDR. In the top-panel of \Fig{clwithstep}, the most striking feature  in $\Delta C_\ell / C_\ell$ is a qualitative transition from rapid oscillations at high $\ell$ to almost no oscillations at low $\ell$. We also note that this transition occurs at higher $\ell$ for models with earlier steps (larger $\zt$). These rapid oscillations are indicative of a shift in the locations of the high-$\ell$ CMB peaks. This is made explicit in the bottom-panel of  \Fig{clwithstep} which shows the shift of the CMB peaks in the model with a step relative to the peak locations in the SIDR reference model. Note that there is no sizeable shift in the position of the low-$\ell$ multipoles, whereas the relative shift to the position of the high-$\ell$ modes increases approximately linearly with $\ell$.

The qualitatively different behavior of high vs. low $\ell$ is due to the different evolution of modes which cross the horizon before and after the step. As we will show below, the locations of the acoustic peaks are sensitive to the expansion rate at the time when the corresponding modes enter the horizon. Low-$\ell$ modes which enter after the step experience the same Hubble rate as the corresponding modes in the reference SIDR model. In contrast, high-$\ell$ modes which enter before the step experience a smaller expansion rate at horizon crossing, which shifts the corresponding peaks towards smaller $\ell$.

It is straightforward to analytically derive the shift of the peak locations for transitions which happen well before matter-radiation equality. Readers less interested in this derivation are encouraged to skip the remainder of this section. The main results are Eqs.~(\ref{eq:dtau}) and (\ref{eq:dgamma}) for the phase shift. Assuming transitions well before equality, we can ignore the contributions of the matter density to both the Hubble parameter and the gravitational potentials during horizon crossing for all modes of interest. For modes that enter either well before or well after the transition, $N(z)$ is a constant at horizon crossing, in which case one can show that the photon number overdensity per coordinate volume $d_\gamma$ is constant before horizon crossing (in conformal Newtonian gauge)~\cite{Bashinsky:2003tk,Baumann:2015rya}. Moreover, the Einstein equation involving the photon-baryon velocity, supplemented with the continuity equation (see, e.g., Eqs.~(23b) and (30) of Ref.~\cite{Ma:1995ey}), yields the initial condition for the conformal-time derivative of $d_\gamma$. The superhorizon initial conditions for the photon perturbation are therefore
\be
\label{eq:initialcondition}
 d_\gamma = -3 \zeta
 ~~,~~ \dot{d}_\gamma = -\frac{1}{2} \, k^2 \, \mathcal{H}^{-1} \, \Phi
 ~,
\ee
where $\zeta$ is the primordial curvature perturbation, $\mathcal{H}$ is the conformal Hubble parameter, and $\Phi$ is the gravitational potential in the notation of Ref.~\cite{Baumann:2015rya}. For low-$k$ modes which enter the horizon well after the step, the initial conditions in \Eq{initialcondition} are therefore identical to the initial conditions in the reference SIDR model. Furthermore, since the WZDR and SIDR models employ identical cosmological parameters, the post-step evolution of the perturbations is also identical. Therefore, there is no shift to the peak locations for such low-$\ell$ modes when comparing these two models. 

On the other hand, large-$k$ modes enter the horizon well before the step, i.e., at a time when the amount of fluid is less than in the reference model. Therefore, these modes experience a slower expansion rate before and during horizon crossing than for the same modes in the SIDR model. After horizon crossing, the gravitational potential decays quickly, effectively suppressing any direct impact that the stepped fluid has on the evolution of the photon perturbations. However, the slower expansion rate encountered by such photon perturbations means that they undergo a larger phase evolution until the step is completed. The resulting phase shift is proportional to the wave number $k$ as well as the difference in conformal time $\Delta \tau$, as defined below in \Eq{dtau}.

To derive $\Delta \tau$ we begin by observing that the perturbation equations are usually written as differential equations with derivatives with respect to conformal time $\tau$ and coefficients which depend on $\mathcal{H}$. To solve the equations one usually begins by substituting $\mathcal{H}=\tau^{-1}$ and integrating. This relation follows from integrating the Friedmann equations, which during radiation domination and for constant $N(z)$ yield 
\begin{align}
	\tau(z) = \mathcal{H}^{-1}(z) + C~,
\end{align}
where $C$ is an integration constant. This integration constant corresponds to the origin of the conformal time and is arbitrary, a reflection of the time translation invariance of the perturbation and Friedmann equations. Making the customary choice $C=0$ then implies $\lim_{z \to \infty} \tau = 0$ and $\mathcal{H}=\tau^{-1}$. We will follow this convention for the reference SIDR model. In order to make the comparison simple, we choose $C$ for the WZDR model such that at late times both models have the same conformal time $\tau$ at fixed $z$.  Using this boundary condition, we find
\begin{align}
	\mathcal{H}^{-1}(z) & = \begin{cases}
		\tau(z) & z \ll z_t \\
		\tau(z) + \Delta\tau & z \gg z_t
		~,
	\end{cases}
\end{align}
in the WZDR model, such that $\Delta\tau$ is the calculable constant
\begin{align}
\label{eq:dtau}
\Delta \tau &= \int_{z_*}^\infty d z \left( \frac{1}{H_{\rm WZDR} (z)}-\frac{1}{H_{\rm SIDR} (z)} \right)
~,
\end{align}
where $H_\text{WZDR, SIDR}$ is the Hubble parameter in the WZDR or SIDR model, respectively. Note that $H_\text{WZDR}=H_\text{SIDR}$ after the step, so that the integrand vanishes for redshifts between recombination at $z_*$ and the step. Matching onto the superhorizon solutions of \Eq{initialcondition}, the well-known solution for the photon perturbation inside the horizon takes the form
\begin{align}
\label{eq:dgamma}
	d_\gamma(z) & \propto \begin{cases}
		\cos[c_\gamma \, k \, \tau(z)] & (k \ll k_t) \\
		\cos[c_\gamma \, k \, (\tau(z) + \Delta\tau)] & (k \gg k_t)
		~,
	\end{cases}
\end{align}
where $c_\gamma \simeq 1/\sqrt{3}$ is the speed of sound of the photon fluid and $k_t$ is defined as the wave number of a mode that enters the horizon during the transition. This solution neglects the drag effect contribution from the gravitational potentials~\cite{Bashinsky:2003tk,Baumann:2015rya}, which modifies both the phase and amplitude of $d_\gamma$ and gives rise to subleading effects. 

From \Eq{dgamma}, we see that the photon perturbations for small-$k$ modes ($k \ll k_t$) are now manifestly identical to those of the reference SIDR model, whereas the high-$k$ modes have a phase shift at recombination approximately given by $c_\gamma k \Delta \tau$. Since low and high $k$ correspond to low and high $\ell$, respectively, we see that \Eq{dgamma} qualitatively reproduces the low and high $\ell$ limits of the phase shift shown in the bottom panel of Fig.~\ref{fig:clwithstep}: low $\ell$ modes have no phase shift whereas high $\ell$ modes have a phase shift given by $\Delta \ell/\ell=\Delta k/k=-\Delta \tau/\tau_*$.
We note that this phase shift can be removed by adjusting $H_0$ to keep the angular size of the sound horizon $\theta_s$ fixed between the two models.\footnote{We have verified numerically in CLASS that the change to $\theta_s$ from the step satisfies $\Delta\theta_s/\theta_s \simeq \Delta\tau /\tau^*$.} However, this amounts to introducing a corresponding phase shift at low-$\ell$. Indeed, the $\ell$-dependent signature of a stepped fluid is distinct from a modification to the angular scale of the sound horizon at recombination.  

The picture described above holds qualitatively even when matter effects are non-negligible at the time of the transition. For small-$k$ modes, which enter the horizon well after the step, the conformal time difference between horizon entry and recombination is the same as in the SIDR model, and thus the small-$k$ modes of both the WZDR and the SIDR models possess the same phase evolution.  On the contrary, large-$k$ modes that enter the horizon well before the step experience a different expansion rate, and thus have more time to evolve between horizon entry until the time of recombination. 

In the next section, we directly confront the data and find that the WZDR model is preferred compared to \lcdm\  and previously studied models of SIDR. In fact, a fit to the CMB and BAO prefers a step location near $\zt \simeq 2 \times 10^4$, the redshift at which multipoles of $\ell \sim 10^3$ reenter the horizon. 

\begin{table}[t!]
\setlength\extrarowheight{3pt}
\centering
\begin{tabular}{|c || c | c | c |} 
\hline
Model& $\Delta \chi^2$ & $N_\text{eff,IR}$ & $H_0$ $(\ksm)$  \\
\hline \hline
$\text{$\Lambda$CDM}$ & 0.0 & $3.04 $ & $67.6 ~\, [67.0,68.3]$ \\  \hline
$\text{$\Lambda$CDM} + \Neff$ & 0.0 & $3.04 ~\, [3.04,3.30]$ & $67.7 ~\, [67.2,69.4]$ \\  \hline
SIDR & $-0.1$ & $3.07 ~\, [3.04,3.43]$ & $67.9 ~\, [67.4, 70.6]$\\ \hline
WZDR & $-1.2$ & $3.27 ~\, [3.05,3.56]$ & $69.1 ~\, [67.6,70.9]$ \\
 \hline
\end{tabular}
\caption{A summary of a fit to the SH0ES-independent dataset $\D$ for various models listed in the first column. The second column gives the $\Delta \chi^2$ of the best-fit compared to  \lcdm. The third and fourth columns display the preferred values of the effective number of neutrinos in free-streaming and strongly-coupled radiation at late times $N_\text{eff,IR}$  (i.e., including the Standard Model neutrino contribution) and the Hubble parameter $H_0$, respectively. The first entry of each cell for $N_\text{eff,IR}$ and $H_0$ gives the best-fit value. The brackets that follow denote the $90\%$ C.L. posterior range, which is defined to be the narrowest interval containing $90\%$ of the integrated posterior density. These are computed directly from the posterior densities and not using Gaussian fits to the posteriors.}
\label{tb:minitab1}
\end{table}

\section{Cosmological Imprints of WZDR}
\label{sec:data}
To understand the implications of this scenario, we perform a fit of various models to a range of cosmological data. We use a modified version of standard CLASS v2.9~\cite{CLASS} (implementing the fluid as described in \App{background}) combined with a MCMC sampler (MontePython v3.4~\cite{Brinckmann:MP}) to study the cosmological constraints on models of interacting radiation. For the WZDR model, we adopt flat priors on the cosmological parameters $ \{\omega_b,\omega_\mathrm{cdm}, \theta_s, n_s,A_s, \tau_\mathrm{reio},\nir, \log_{10} z_t \}$ with the following ranges $\log_{10} \zt \ \in [4.0, 4.6]$ and $ \nir \geq 0$.\footnote{We restrict the range of $z_t$ in order to avoid very late and very early transitions because with current data such points are indistinguishable from conventional SIDR models.} We assume a Standard Model neutrino sector that undergoes a standard cosmological history, consisting of one massive neutrino species with a mass of $0.06 \ \eV$ and two massless neutrino species.

As in Ref.~\cite{Schoneberg:2021qvd}, we consider the following datasets in our default analysis. The dataset $\D$ includes the Planck 2018 dataset~\cite{Planck:2018vyg}, including TT,TE, and EE in both low ($\ell < 30$) and high ($\ell \geq 30$) multipole ranges, as well as CMB lensing~\cite{Planck:2018vyg} and the full set of nuisance parameters. It also includes late-universe constraints, such as BAO measurements from BOSS DR12 ($z=0.38,0.51,0.61$)~\cite{BOSS} and small $z$ measurements from 6dF ($z=0.106$)~\cite{6dF} and MGS ($z=0.15$)~\cite{MGS} catalogs, as well as  PANTHEON supernovae data~\cite{Scolnic:2017caz}. In addition to our baseline dataset $\cal D$, we also consider a dataset $\cal D+$ that additionally includes measurements from SH0ES~\cite{Riess:2020fzl} via a prior on the intrinsic magnitude of supernovae $M_b$. We summarize our results for a range of models fitting to $\cal D$ and $\cal D+$ in \App{triangles}.

When we consider the impact of WZDR on cosmological observables, the immediate question is: relative to what? Because the presence of new interacting radiation changes integrated quantities (such as the sound horizon at recombination), the change in any one parameter naturally changes others, making direct comparisons between different models complicated. Two natural points of comparison are to models with additional strongly-coupled  or free-streaming radiation  \textit{without} a step. The former is the conventional SIDR model discussed above, and the latter is a minimal extension of \lcdm, in which the effective number of free-streaming neutrino species is allowed to take values greater than the \lcdm\ contribution from the Standard Model neutrino sector, i.e., $\Neff > 3.044$ (denoted as $\Lambda \text{CDM} + \Neff$)~\cite{Froustey:2020mcq,Bennett:2019ewm,Akita:2020szl}. Using  \Tab{minitab1}, we proceed to make these comparisons, which shows results of a fit to the SH0ES-independent dataset $\D$. For convenience, we define the variable $N_\text{eff,IR}$ as the late-time value of the effective number of neutrinos in free-streaming and strongly-coupled radiation (including the Standard Model neutrino contribution).

\begin{table}[t!]
	\setlength\extrarowheight{3pt}
	\centering
	\begin{tabular}{|c || c | c | c |} 
		\hline
		Model& $\Delta \chi^2$ & $N_\text{eff,IR}$ & $H_0$ $(\ksm)$  \\
		\hline \hline
		$\text{$\Lambda$CDM}$ & 0.0 & $3.04$ & $68.2 ~\, [67.5,68.9]$ \\  \hline
		$\text{$\Lambda$CDM} + \Neff$ & $-5.7$ & $3.37 ~\, [3.20,3.63]$ & $70.0 ~\, [68.9,71.6]$ \\  \hline
		SIDR & $-10.6$ & $3.51 ~\, [3.31,3.77]$ & $71.0 ~\, [69.6, 72.6]$\\ \hline
		WZDR & $-15.1$ & $3.63 ~\, [3.37,3.92]$ & $71.4 ~\, [69.7,73.0]$ \\
		\hline
	\end{tabular}
	\caption{As in \Tab{minitab1}, but after fitting to the dataset $\D+$ that includes SH0ES.}
	\label{tb:minitab1a}
\end{table}

\begin{figure*}[t]
\centering
\includegraphics[width=0.44\textwidth]{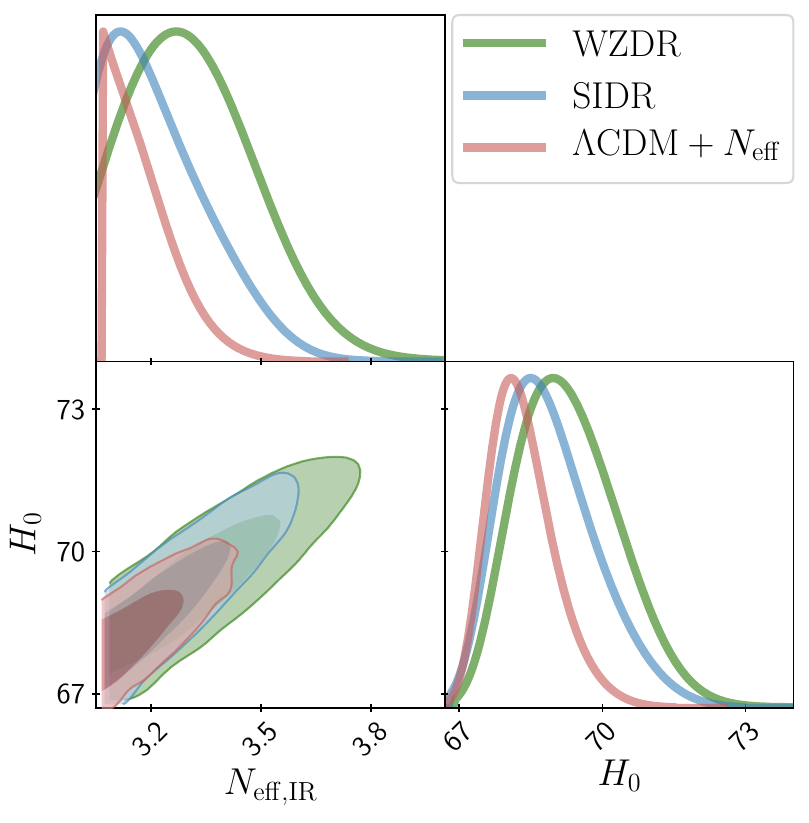}\hskip 0.1in
\includegraphics[width=0.44\textwidth]{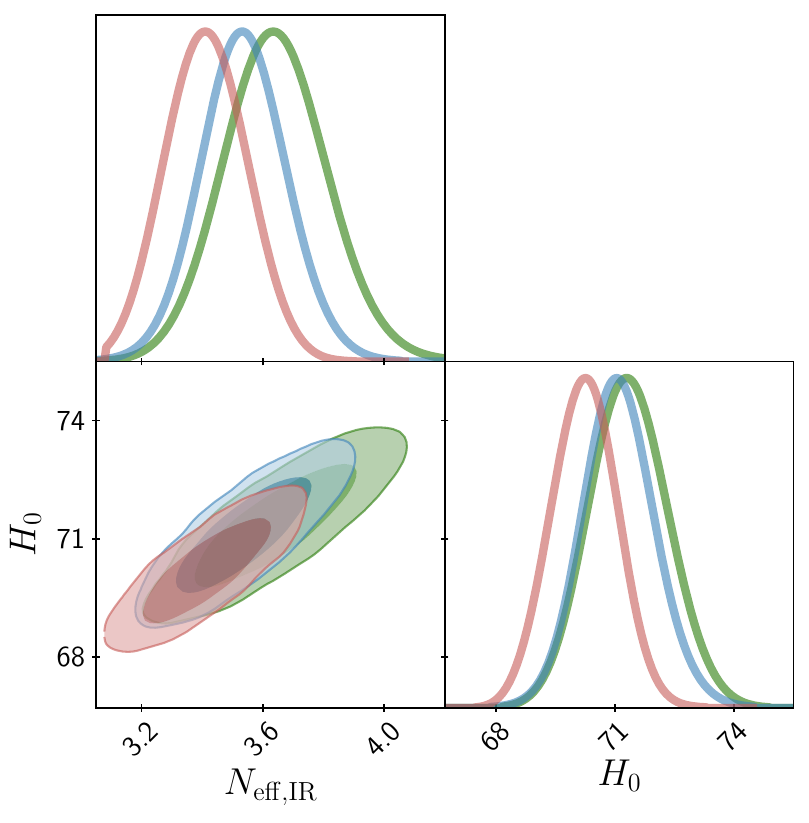}   \caption{Comparison of the marginalized 1D and 2D posterior distributions for the Hubble parameter $H_0$ and the late-time value of the effective number of neutrinos in radiation $N_\text{eff,IR}$ (including the Standard Model neutrino contribution) for the $\Lambda \text{CDM} + \Neff$, SIDR, and WZDR models when fitting to the dataset $\D$ (not including SH0ES) in the left set of panels or $\D+$ (including SH0ES) in the right set of panels.}
\label{fig:h0plots}
\end{figure*}

We note at the outset that the best-fit points for the dataset $\D$ in conventional models involving additional free-streaming or strongly-interacting radiation are either identical or very close to \lcdm\ and do not improve the $\chi^2$ by more than $\order{0.1}$. This changes with the addition of the WZDR-step. Considering only data from $\D$, we see a slight improvement going from SIDR to WZDR, reducing the best-fit $\chi^2$ by $1.2$, about as would be expected with the addition of the one new parameter $\zt$, which has a posterior mean-value of $\zt = 1.9_{-0.6}^{+0.9} \times 10^4$. Thus, although the quantitative pull is small, there is a nonzero preference in the data, independent of SH0ES, for a step in an additional interacting fluid component.\footnote{Many previous analyses have assumed that the additional radiation was also present during BBN, increasing the predicted abundance of primordial helium $Y_p$. Since that is a much earlier era, we instead assume that the radiation is populated well after BBN, at temperatures below $\sim 100 \ \keV$~\cite{Berlin:2019pbq}. This is a natural assumption in the context of interacting fluids, and we make this same assumption for the SIDR model, so as not to penalize it compared to WZDR.
For completeness, we also show results for SIDR and WZDR in which the energy density in interacting radiation is present during BBN in \Tab{fitresults} of \App{triangles}. For the $\Lambda \text{CDM} + \Neff$ model, we assume that the extra radiation is present during BBN, as considered in Ref.~\cite{Schoneberg:2021qvd}.} 

Remarkably, the preferred value for $H_0$ in a fit to $\D$ is shifted to larger values in the WZDR model, with a best-fit of $H_0= 69.1 \ \ksm$, as compared to $H_0 = 67.9 \ \ksm$ for SIDR, $H_0=67.7 \ \ksm$ for $\text{$\Lambda$CDM} + \Neff$, and $H_0=67.6 \ \ksm$ for \lcdm.
A similar trend is evident in the 90\% C.L. ranges listed in \Tab{minitab1}. This immediately raises the question: can WZDR help address the existing tension between the \lcdm-inferred value of $H_0$ and the late-universe measurement of $H_0=(73.2 \pm 1.3) \ \ksm$ by the SH0ES collaboration?

The short answer to this question is: yes, the presence of the WZDR-step does allow for a significant reduction to this tension. Earlier analyses have shown that additional free-streaming radiation can naturally allow for a larger value of $H_0$ when fitting to the dataset $\D+$ that includes SH0ES, but only at the cost of significantly worsening the fit to the SH0ES-independent dataset $\D$. Making this radiation interacting (as in the SIDR model) somewhat ameliorates the issue, but this is still constrained by the high-$\ell$ multipoles of the CMB polarization power spectrum~\cite{Bernal:2016gxb,Blinov:2020hmc,Brinckmann:2020bcn}. In the WZDR model, this is compensated by the the $\ell$-dependent modifications to the CMB, allowing for additional levels of interacting radiation.

We provide additional results of our dedicated MCMC analysis in \Fig{h0plots}, which shows the posteriors for $H_0$ and the late-time value of the effective number of neutrinos in free-streaming and strongly-coupled radiation for the $\text{$\Lambda$CDM}+ \Neff$, SIDR, and WZDR models (the full set of posteriors for each of these scenarios is provided in \App{triangles}). As for conventional early-universe solutions to the Hubble tension, additional radiation is correlated with larger values of $H_0$, corresponding to an approximately fixed angular size of the sound horizon at recombination~\cite{Hou:2011ec}. Most notable in  \Fig{h0plots} is the fact that WZDR predicts $H_0$ and $N_\text{eff,IR}$ posteriors that extend out to larger values, in the case that SH0ES is either included (right panel) or not included (left panel) in the analysis. We show the resulting best-fit values and posterior ranges for the full $\D+$ dataset in \Tab{minitab1a}.

But, of course, simply predicting a larger value of $H_0$ is not a solution to the tension if it simply provides an overall bad fit to the data. Recently, Ref.~\cite{Schoneberg:2021qvd} established a rubric for comparing models that could address this tension with three basic measures: GT (Gaussian tension), $\QDMAP$ (difference of the maximum a posteriori), and $\Delta \text{AIC}$ (Akaike information criterium). The values of these measures are provided in \Tab{minitab2} for \lcdm \, and the three benchmark radiation models. We now briefly summarize each of these in turn (see Refs.~\cite{Schoneberg:2021qvd,Raveri:2018wln} for further details). 

\begin{table}[t]
\centering
\begin{tabular}{|c || c | c | c |} 
\hline
Model& GT & $\QDMAP$ &  $\Delta \text{AIC}$  \\
\hline \hline
$\text{$\Lambda$CDM}$ & $4.5 \sigma$ & $4.5 \sigma$ & $0.0$\\  \hline
$\text{$\Lambda$CDM} + \Neff$ & $3.7 \sigma$ & $3.8 \sigma$ & $-3.7$ \\  \hline
SIDR & $3.1 \sigma$ & $3.1 \sigma$ & $-8.6$ \\ \hline
WZDR & $2.7 \sigma$ & $2.4 \sigma$ & $-11.1$ \\
 \hline
\end{tabular}
\caption{The GT (Gaussian tension), $\QDMAP$ (difference of the maximum a posteriori), and $\Delta \text{AIC}$ (Akaike information criterium) measures, which together serve to address the success in resolving the Hubble tension for various models.}
\label{tb:minitab2}
\end{table}

GT and $\QDMAP$ both address the residual level of tension between the results of SH0ES (expressed as the measured supernova magnitudes $M_b$) and a model-predicted value obtained by fitting to the SH0ES-independent dataset $\D$. In particular, GT is defined as the difference in the $M_b$ mean-value as determined from either method, in units of $\sigma$ (the standard deviation obtained from $\D$ and SH0ES added in quadrature). $\QDMAP$, on the other hand, is defined as the square root of the difference between the best-fit $\chi^2$ obtained by either fitting a model to $\D+$ or $\D$, in units of $\sigma$. Smaller values of either measure signify a reduced tension in the data, and for Gaussian distributed posteriors, they are equivalent. We find that while $\text{$\Lambda$CDM} + \Neff$ and SIDR reduce the $4.5 \sigma$ tension between \lcdm\ and SH0ES to the $\sim 4 \sigma$, $3 \sigma$ level, respectively, WZDR significantly reduces it further still, down to $\sim 2.5 \sigma$.

$\Delta \text{AIC}$ measures the success in fitting to the full dataset including SH0ES, defined as the difference in the best-fit $\chi^2$ for $\D+$ between a given model and \lcdm\ with a $\chi^2$ penalty of $+2$ for each additional (beyond \lcdm) model parameter:
\be
\Delta \text{AIC} =\chi^2-\chi_{\Lambda \rm CDM}^2+2 \times (\rm new\ parameters)
\, .
\ee
Hence, more negative values signify increased agreement with the full dataset. As shown in \Tab{minitab2}, this measure further illustrates the dramatic improvement in a fit involving WZDR. In particular, while $\Lambda \text{CDM} + \Neff$ and SIDR reduce the $\Delta \text{AIC}$ down to $-3.7$ and $-8.6$, respectively, this is improved further to $-11.1$ in the WZDR model (corresponding to a best-fit $\chi^2$ which is lower by $-15.1$ compared to \lcdm). 

With the addition of a single parameter, WZDR provides an improvement of $\Delta \chi_{\D+}^2 = -4.5$ when fitting to $\D+$, as compared to SIDR (see \App{triangles}). Also noteworthy is the fact that when fitting to the dataset $\D+$ including SH0ES, the fit to the SH0ES-independent dataset $\D$ for WZDR is comparable to that of \lcdm. In contrast, both \lcdm+$\Neff$ and SIDR provide markedly worse fits to $\D$ when fitting to the full $\D+$ (see \Tab{bestfitDp}).

To summarize, the WZDR model exemplifies a significant improvement in each of the three standardized measures of Ref.~\cite{Schoneberg:2021qvd}, as shown in \Tab{minitab2}. Thus, this model provides a potential scenario to explain this discrepancy. Moreover, as future data at high-$\ell$ arrive, such as from the Simons Observatory or CMB-S4, a direct comparison to the CMB, even without SH0ES data, should further clarify the degree of this success.

\section{Generalizations and Stepped Dark Sectors}
\label{sec:4param}

Although the minimal supersymmetric model for interacting radiation (WZDR) has two free parameters ($\nir$ and $z_t$), one can envision generalizing this to a scenario where the parameter $N_\text{IR} / N_\text{UV}$ is not a fixed number, as in \Eq{step}, for instance if $\psi$ is a fundamental and $\phi$ an adjoint under a global symmetry rather than residing in a single chiral superfield. This then makes $N_\text{IR} / N_\text{UV}$ an adjustable parameter of the model. Additionally, if the dark radiation is populated via late-time freeze-in from neutrinos, the energy density in free-streaming radiation $\nfs$ can be modified~\cite{Berlin:2017ftj, Berlin:2018ztp}. We thus take $\nfs$ as a free parameter, as well.

Thus from a physical point of view, in such models the four free model parameters are the effective number of neutrinos in interacting species $\nuv$ at early times, the number of interacting species  $\nir$ at late times, the redshift at which the dark sector drops below a mass-threshold $z_t$, and the amount of free-streaming radiation $\nfs$. Such a parameterization is a simple extension of interacting fluid models, motivated by the simple presence of a mass-threshold in the sector, which is essentially required in models involving scalars. We refer to this generalization of a stepped fluid as ``General StepDR."

We perform a fit of the General StepDR model to the datasets  $\D$ and $\D+$.  The parameters for this model are: $\{ \omega_b,\omega_\mathrm{cdm}, \theta_s, n_s,A_s, \tau_\mathrm{reio},\nuv, \nir, \nfs, \log_{10} \zt \} $ with the ranges  $\log_{10} \zt \ \in [4.0, 4.6]$, $ \nuv \geq 0 $, and $ \nir \geq \nuv $. Additionally, we assume one massive neutrino species with a mass of $0.06 \ \eV$. The results of this analysis are included in \App{triangles}. From the full set of posterior distributions in \App{triangles}, we see that the introduction of these additional parameters widens the posterior density for $\nuv$ and $\nir$, while the favored region of parameter space for the General StepDR model is consistent with the favored region in the restricted WZDR model. We also see from \App{triangles} that the best-fit $\chi^2$ values are reduced by $\order{1}$ when comparing the General StepDR and WZDR models, as expected from the introduction of two additional free parameters.

\section{Discussion}
\label{sec:conclusion}
Models of interacting radiation provide a plausible and testable extension beyond \lcdm. In many such scenarios there is a natural mass-threshold present, corresponding to the scale of the force-carrier. In such models, there is a ``step'' as the massive mediator entropy is transferred to the remaining degrees of freedom. The simplest such model - a single fermion and a single scalar - is nicely packaged into a well-known supersymmetric theory, namely the Wess-Zumino model. Such a framework naturally allows for a scalar mass which is light enough to be relevant during the era in which the CMB-detectable modes are inside the horizon.

We emphasize that this framework of stepped dark radiation is a theoretically simple one. The Standard Model has copious mass-thresholds, including the critical late threshold of the electron mass where photons are heated in analogous fashion. Our purpose here is to understand the implications of such a step in the context of present-day cosmological data. There is a well-known tension between CMB data and late-universe measurements of the Hubble constant, and so it is important to distinguish them.

Remarkably, WZDR provides a good fit to the CMB data, alone. The best-fit point, not including SH0ES data, prefers additional radiation $\nir = 0.23$, excluding $\nir=0$ at better than 90\% confidence (Table \ref{tb:minitab1}). This contrasts conventional interacting radiation which has a best-fit of $\nir=0.03$, and free-streaming radiation which has a best-fit of $\Delta \Neff = 0$. At the same time, the WZDR model prefers a higher value of $H_0 = 69.1 \ \ksm$ with a wide range of allowed values at 90\% confidence, $[67.6,70.9] \ \ksm$. Including SH0ES in the analysis changes the values somewhat, but qualitatively, the picture does not shift, preferring a best-fit value $H_0 = 71.4  \ \ksm$. 

Comparing WZDR to SIDR with the same late-time parameters, the change to the CMB power spectrum is an $\ell$-dependent phase shift.
Modes entering the horizon earlier, with relatively less radiation present, experience a larger sound horizon, thus shifting the peaks to lower $\ell$ multipoles. While our analysis does not analyze $\chi^2$ values for separate $\ell$ regions (see, e.g., Ref.~\cite{Hill:2021yec}, $\ell < 650$), the WZDR and General StepDR models give a good fit to all data, including high-$\ell$ data.

We should also comment on another discordance in cosmological data, the $S_8$ tension, which we have so far ignored. The \lcdm-preferred value of $S_8$, as  inferred from fitting to the CMB, is about $2\sigma$ higher than the values inferred from weak lensing surveys~\cite{Hildebrandt:2016iqg,Hildebrandt:2018yau,DES:2017myr,HSC:2018mrq}. This tension motivates new physics that suppresses the matter power spectrum. We have not included such large scale structure (LSS) data in our fits, in part because the $S_8$ tension is not as severe or robust as the $H_0$ tension,  LSS data was not included in the fits of Ref.~\cite{Schoneberg:2021qvd}, and the dark radiation models that we have investigated do not significantly alter the prediction for $S_8$ compared to \lcdm\  (see \App{triangles}), and so neither improve nor worsen the fit to LSS data.

However, we remark that in the context of an interacting sector of dark radiation, it is very natural to expect a component of the dark matter to couple to dark radiation. Such a coupling endows the dark matter with pressure that slows the growth of structure and therefore predicts a lower value for $S_8$~\cite{Buen-Abad:2015ova,Lesgourgues:2015wza,Buen-Abad:2017gxg,Bansal:2021dfh}. This pressure depends on a new parameter: the coupling between dark matter and dark radiation. Thus, it would be straightforward to resolve the $S_8$ tension in a natural extension of the stepped-fluid models considered here.  Such a model could arise by including an additional field $Y$ with a $Z_2$ symmetry and superpotenial term of the form 
$X Y^2$. This would naturally yield an interacting dark matter subcomponent. 

In our analysis, we have assumed the standard BBN-predicted value of primordial helium, assuming the dark radiation is not present at that time. It's quite easy for the radiation to appear at late times, for instance through the decay of a relic particle, for which the allowed value of $\nuv$ can be anything. If the dark radiation was in thermal contact at early times such that $\nuv \sim \order{0.1}$ at BBN, it can be boosted by a factor of a few if the dark sector has  a strongly coupled gauge group which confines with a mass gap, such as supersymmetric QCD. One might also study this scenario assuming the radiation is present at BBN, naturally raising the level of primordial helium, and likely worsening the fit to a small degree. 

It is also worth comparing WZDR and General StepDR to other models that invoke qualitatively similar dynamics. For instance, Ref.~\cite{Bansal:2021dfh} recently discussed a mirror copy of the Standard Model within the context of the Hubble tension. Although such theories involve some of the same ingredients as discussed in our work, they also involve additional dynamics arising from transitions from strongly-coupled to free-streaming radiation and epochs in which the dark sector undergoes its own version of nucleosynthesis and recombination. However, we are unable to meaningfully compare to the fits of Ref.~\cite{Bansal:2021dfh}, since these were performed only with data that include SH0ES and LSS. We also note that of all the models catalogued in Ref.~\cite{Schoneberg:2021qvd}, the one involving a neutrino-coupled majoron~\cite{Escudero:2019gvw,EscuderoAbenza:2020egd,Escudero:2021rfi} is the most similar to WZDR. However, WZDR is much more successful in addressing the $H_0$ tension, as quantified by comparing the results of \Tab{minitab2} and Ref.~\cite{Schoneberg:2021qvd}.

Going forward, as data improve, there is a great hope that the CMB data alone will give clarity on whether a scenario similar to ours provides a reasonable description of Nature. A recent analysis by Ref.~\cite{Hill:2021yec} showed that an Early Dark Energy model could be distinguished with future ACT data at the level of $25\sigma$ from the current best-fit \lcdm\ point. The WZDR model has similar sized residuals, and thus we can be hopeful that these data might distinguish it from \lcdm\ as well. A proper analysis is warranted, however.

In summary, we have studied a simple, natural, theoretical model extension beyond \lcdm, namely interacting radiation with a step, arising from a straightforward mass-threshold. This minor extension provides a good fit to the CMB data with a higher value of $H_0$. Including SH0ES data strengthens the evidence for additional radiation, as well as the step, improving the $\chi^2$ by 15.1 compared to \lcdm\ and by 4.5 compared to conventional SIDR, with only a single additional parameter (the step location). As further data appear, we may learn the Hubble tension is the first indication of a dynamic and interacting dark sector.

\section*{Acknowledgements}
We thank Vivian Poulin, Nils Sch\"{o}neberg, and Andrea P\'{e}rez S\'{a}nchez for helpful discussions and comparing data points from the ``$H_0$ Olympics"~\cite{Schoneberg:2021qvd}. The work of D.A., M.J., and M.S. is supported by the U.S. Department of Energy (DOE) under Award DE-SC0015845. A.B. is supported by the James Arthur Fellowship. N.W. is supported by NSF under award PHY-1915409, by the BSF under grant 2018140, and by the Simons Foundation. Our MCMC runs were performed on the Shared Computing Cluster, which is administered by Boston University's Research Computing Services.

\bibliographystyle{utphys}
\bibliography{hostep}
\appendix
\section{Evolution of a Stepped Fluid}
\label{app:background}

We assume that the interaction rate in the dark fluid is large enough such that local thermal equilibrium is maintained. At the background (homogeneous) level, this implies that entropy conservation is sufficient to track the evolution of the fluid parameters 
\bea\label{eq:define N,w,cs2}
N\equiv \frac{\rho}{\rho^{1\nu}}\, , \quad w\equiv \frac{p}{\rho} \, , \quad
c_s^2\equiv \frac{dp/dT_d}{d\rho/dT_d}
\, ,
\eea
through the step, where $\rho^{1 \nu}$ is the energy density of a single neutrino in \lcdm. $N$, $w$, $c_s^2$ are functions of the dark sector temperature $T_d$, which depends on the scale factor $a$. Determining $N(a)$, $w(a)$ and $c_s^2(a)$ is the purpose of this Appendix. 

The dark energy density $\rho(T_d)$ and pressure $p(T_d)$ of the interacting dark fluid have contributions from both massive and massless particles. For a single particle of mass $m$ with $g$ internal degrees of freedom
and temperature $T_d$ in equilibrium, the energy density and pressure are
\begin{align}
\rho(T_d)&=g \int \frac{d^3p}{(2 \pi)^3} \sqrt{p^2+m^2} \frac{1}{e^{\sqrt{p^2+m^2}/T_d}\pm1}
\nonumber \\
p(T_d)&=\frac{g}{3} \int \frac{d^3p}{(2 \pi)^3} \frac{p^2}{\sqrt{p^2+m^2}} \frac{1}{e^{\sqrt{p^2+m^2}/T_d}\pm1} \, ,
\end{align}
where $\pm 1$ corresponds to fermions/bosons, respectively. It will be convenient to rewrite this by factoring out the energy density and pressure of a single relativistic boson degree of freedom, $\rho_B(T_d)=3 p_B(T_d)=T_d^4 \pi^2/30$. This gives
\begin{align}
\rho(T_d)&\equiv g_*\, \rho_B(T_d)\, \hat\rho(x)
\nonumber \\
p(T_d)&\equiv g_*\, p_B(T_d)\, \hat p(x)
\, ,
\end{align}
where $x=m/T_d$, and we have defined the effective degrees of freedom, $g_*=g$ for bosons and $g_* = (7/8)\, g$ for fermions.
Note that the dimensionless integrals $\hat \rho$ and $\hat p$ depend on the temperature and mass only through their ratio $x$. 
$\hat\rho(0)=\hat p(0)=1$ and both decay exponentially for large $x$. $\hat \rho$ and $\hat p$ are quite similar for bosons and fermions and we approximate them by dropping the $\pm1$ terms in the distribution functions. This approximation works to better than 8\% for all $x$, and to better than 4\% for $x$ for which $\hat \rho$ and $\hat p$ are not exponentially suppressed and therefore irrelevant.

The (Maxwell-Boltzmann) integrals can now be evaluated analytically 
\begin{align}
\label{eq:rhohat}
\hat\rho(x) &= \int_0^\infty\!\!\!\! dq\, q^2 \sqrt{q^2+x^2}\ e^{-\sqrt{q^2+x^2}} \bigg/ \int_0^\infty\!dq\, q^3 \ e^{-q}
\nonumber \\
&= \frac{x^2}{2} K_2(x)+\frac{x^3}{6} K_1(x)
\nonumber \\
\hat p(x) &=  \int_0^\infty\!\!\!\! dq \, \frac{q^4}{\sqrt{q^2+x^2}}\ e^{-\sqrt{q^2+x^2}} \bigg/  \int_0^\infty\!\!\!\! dq \, q^3\, e^{-q}
\nonumber \\
&= \frac{x^2}{2} K_2(x)
\, ,
\end{align}
in terms of modified Bessel functions of the second kind, $K_i$.

The energy density of a fluid which contains $g_*^\text{IR}$ massless and $g_*^\text{UV}-g_*^\text{IR}$ massive particles can then be written as\footnote{For simplicity, we assume a single mass scale $m$. It is straightforward to generalize to several particles with different masses.}
\bea
\label{eq:apprho}
\rho(T_d) 
&=& g_*^\text{IR} \rho_B(T_d) (1+r_g \hat\rho(x))
\, ,
\eea
where we defined the step size $r_g\equiv (g_*^\text{UV}-g_*^\text{IR})/g_*^\text{IR}$. The pressure is
\bea
\label{eq:appp}
p(T_d)&=& g_*^\text{IR} p_B(T_d) (1+r_g\hat p(x)) \ .
\eea
Eqs.~(\ref{eq:apprho}) and (\ref{eq:appp}) again include a factor of $7/8$ in the definition of $g_*^\text{IR}$ and $g_*^\text{UV}$ for each fermion. Therefore, the equation of state is
\bea
\label{eq:w}
w(x)=
\frac13 \frac{1+r_g\hat p(x)}{1+r_g\hat \rho(x)} 
=\frac13 -\frac{r_g}{3}\, \frac{\hat\rho(x) - \hat p(x)}{1+r_g\hat \rho(x)} 
\, ,
\eea
and the speed of sound is
\begin{align}
\label{eq:cs2}
c_s^2(x)&=\frac{dp/dx}{d\rho/dx}
=\frac13 \frac{1+r_g (\hat p(x) -\frac{x}{4} \hat p'(x))}{1+r_g (\hat \rho(x) -\frac{x}{4} \hat \rho'(x))}
\nonumber \\
&=\frac13-\frac{r_g}{36}\, \frac{x^2 \hat p(x)}{1+r_g (\frac34\hat \rho(x) +(\frac14+\frac{x^2}{12})\hat p(x))}
\, ,
\end{align}
where in the second line we used Bessel function recursion relations to rewrite $\hat \rho'$ and $\hat p'$ in terms of $\hat \rho$ and $\hat p$. We see that both $w$ and $c_s^2$ are 1/3 minus a term proportional to the step size $r_g$ which vanishes for both small and large $x$ (see \Fig{csofa}).

The expressions above are written in terms of $x=m/T_d$, but we will need them as functions of scale factor $a$. Therefore, we determine $x(a)$ from entropy conservation in the dark fluid
\bea
\label{eq:entropy}
S\propto a^3 \frac{\rho(T_d)+p(T_d)}{T_d} = a_0^3 \frac{\rho(T_{d0})+p(T_{d0})}{T_{d0}} 
\, ,
\eea
where $a_0=1$ is the scale factor today and $T_{d0}$ is the dark sector temperature today. Substituting the expressions for $\rho$ and $p$ from Eqs.~(\ref{eq:apprho}) and (\ref{eq:appp}), assuming that the massive particles have annihilated away today so that $\hat \rho(m/T_{d0})=\hat p(m/T_{d0})=0$, and defining the transition scale factor $a_t\equiv T_{d0}/m$, we obtain
\bea
\label{eq:aofx}
\left(\frac{x a_t}{a}\right)^3=1+\frac{r_g}{4}(3 \hat\rho(x)+\hat p(x))
\, ,
\eea
which can be solved numerically for $x(a)$ for any given set of model parameters $r_g$ and $a_t$.
Note that well after the transition, $\hat\rho(x)$ and $ \hat p(x) $ are negligibly small. In this case, Eq.~(\ref{eq:aofx}) simplifies to $x a_t= a$, or equivalently $T_d=T_{d0}/a$. At early times, well before the transition, $x\ll1$ so that $\hat\rho(0)=\hat p(0)=1$ and therefore $x a_t= (1+r_g)^{1/3}a$ or $T_d=(1+r_g)^{-1/3}T_{d0}/a$. Thus, the temperature of the dark sector in the UV is smaller than the naive $1/a$ scaling by a factor of $(1+r_g)^{-1/3}$ because of the step.

Finally, we can calculate the ``effective number of neutrino species''  from Eq.~\eqref{eq:define N,w,cs2}
\bea
\label{eq:N}
N(x) &=& \nir \frac{1+r_g\hat\rho(x)}{(1+r_g(\frac34 \hat\rho(x)+\frac14\hat p(x)))^{4/3}} 
\, ,
\eea
where we used
$\rho^{1\nu}=\frac74\frac{\pi^2}{30} (\frac{T_{\nu0}}{a})^4$, substituted Eq.~(\ref{eq:aofx}), and identified
\bea
\nir &=& \frac{g_*^\text{IR}}{7/4} \left(\frac{T_{d0}}{T_{\nu0}}\right)^4 
\, ,
\eea
by taking the IR ($x\rightarrow \infty$) limit.
Taking the UV ($x\rightarrow 0$) limit gives the UV endpoint of the step
\bea
\nuv&=& \frac{\nir}{(1+r_g)^{1/3}} = \nir \left(\frac{g_*^\text{UV}}{g_*^\text{IR}}\right)^{1/3}
\, .
\eea
This equation confirms Eq.~(\ref{eq:step}) and relates the parameter $r_g$ to $\nir/\nuv$ by
\bea
\label{eq:rg}
r_g=\left(\frac{\nir}{\nuv}\right)^3-1 \ .
\eea

In summary, the background quantities describing the fluid are $w(x)$, $c_s^2(x)$, and $N(x)$, which are given in Eqs.~(\ref{eq:w}), (\ref{eq:cs2}), and (\ref{eq:N}). $x(a)$ is obtained by numerically solving (\ref{eq:aofx}), $r_g$ is given in (\ref{eq:rg}), and $\hat \rho(x)$ and $\hat p(x)$ in (\ref{eq:rhohat}). These are the equations we implement in CLASS, along with the standard perturbation equations for a strongly-coupled fluid (in synchronous gauge)~\cite{Ma:1995ey}
\bea
 \label{eq:perteq}
 \dot{\delta} &=& - \left( 1 + w  \right) \left( \theta + \frac{\dot{h}}{2} \right) - 3  \mathcal{H} \left( c_s^2 - w  \right) \delta \nonumber \\
  \dot{\theta} &=& \frac{k^2 c_s^2}{1 + w} \delta -  \mathcal{H} \left(1 - 3 c_s^2 \right) \theta
 ~.
\eea

The parameters of the stepped dark fluid are the transition scale $a_t$ (or equivalently, $z_t = 1 / a_t - 1$), the amount of dark radiation in the IR, $\nir$, and the ratio $r_g$ determining the size of the step (or equivalently, $\nir/\nuv$).  In the WZDR model, $r_g=8/7$.

\onecolumngrid
\section{Triangle Plots and Parameter Values}
\label{app:triangles}

\begin{figure*}[!htbp]
	\centering
	\includegraphics[width=1.\textwidth]{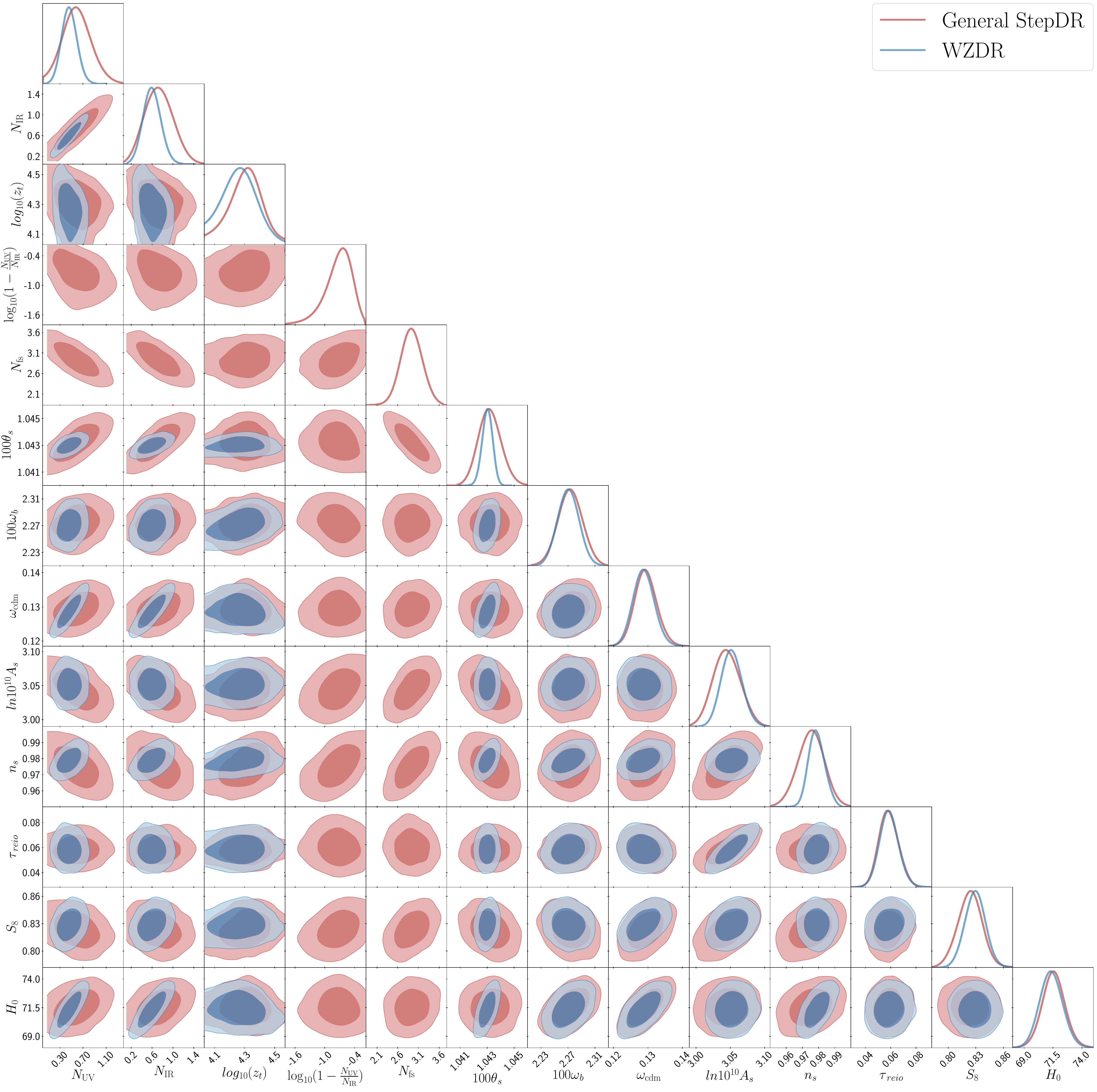}
	\caption{A comparison of the posteriors for the minimal stepped fluid (WZDR in blue) and generalized stepped fluid (General StepDR in red) when fitting to the dataset $\D+$ that includes SH0ES. The dark and light shaded regions correspond to $68.3\%$ and $95.4\%$ C.L., respectively. It is interesting to note that while the introduction of two additional free parameters in the General StepDR model widens the posterior density for $\nuv$ and $\nir$, the favored region of parameter space is approximately unchanged compared to the restricted WZDR model.}
	\label{fig:wzdr-4param-triangle}
\end{figure*}

\begin{figure*}[!htbp]
	\centering
	\includegraphics[width=1.\textwidth]{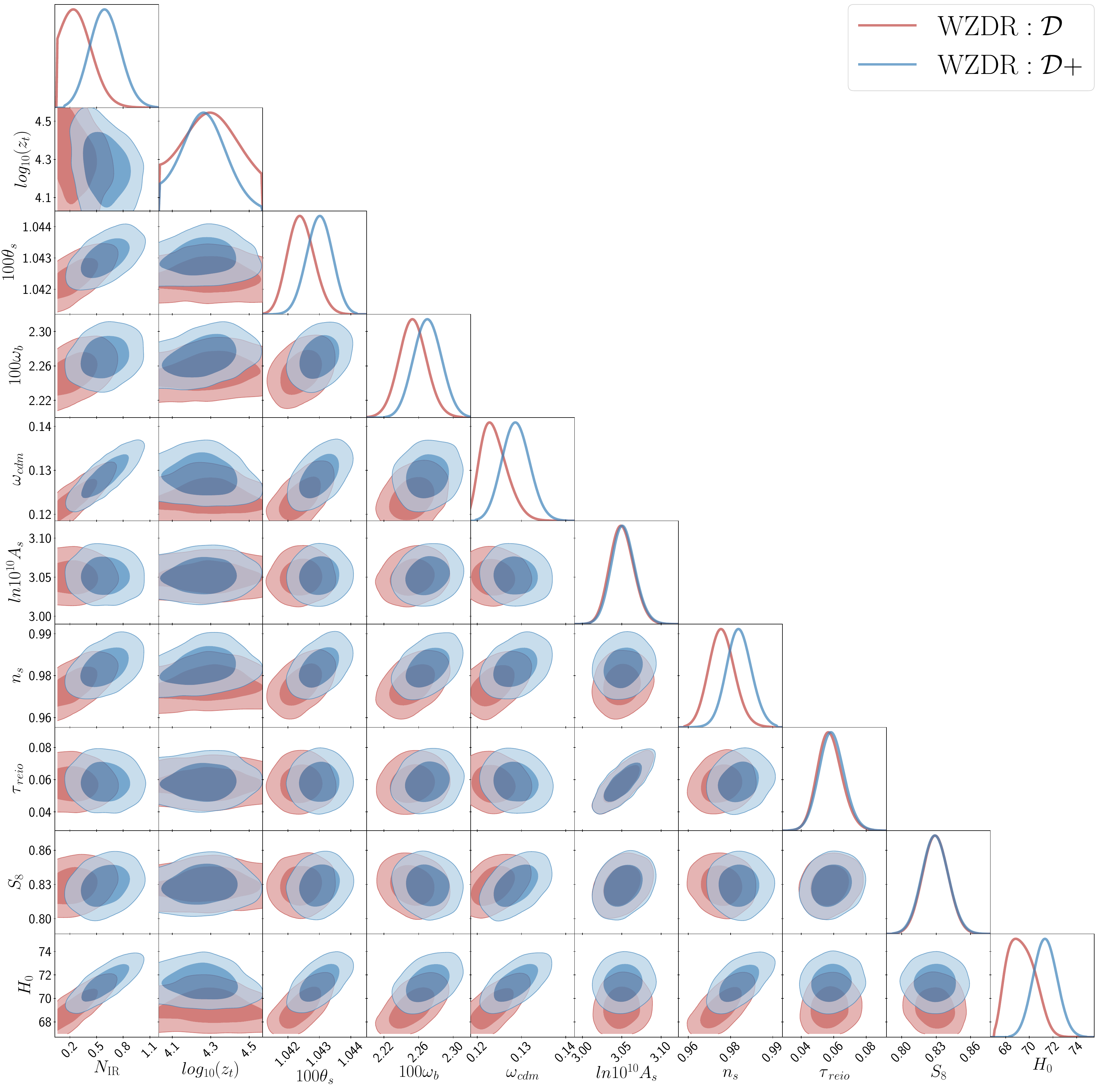}
	\caption{A comparison of the posteriors for the minimal stepped fluid (WZDR) when fitting to either the SH0ES-independent dataset $\D$ (red) or the dataset $\D+$ that includes SH0ES (blue). The dark and light shaded regions correspond to $68.3\%$ and $95.4\%$ C.L., respectively. Comparing the two sets of posteriors, we see that a fit to data including SH0ES prefers a larger value of $H_0$ and a correspondingly larger energy density in dark radiation, as expected, while leaving the preferred location of the transition $z_t$ nearly unchanged. Since an enhanced radiation density strengthens the effect of the step on the CMB, the data is increasingly sensitive to the location of the transition, as can be noted from the narrower posterior for $z_t$ when fitting to $\D+$. Along with the increase in radiation density, the fit to $\D+$ also requires increases in $\omega_{\rm cdm}$, $\omega_b$, as well as $\theta_s$ and $n_s$.}
	\label{fig:wzdr-SH0ES-noSH0ES-triangle}
\end{figure*}

\begin{figure*}[!htbp]
	\centering
	\includegraphics[width=1.\textwidth]{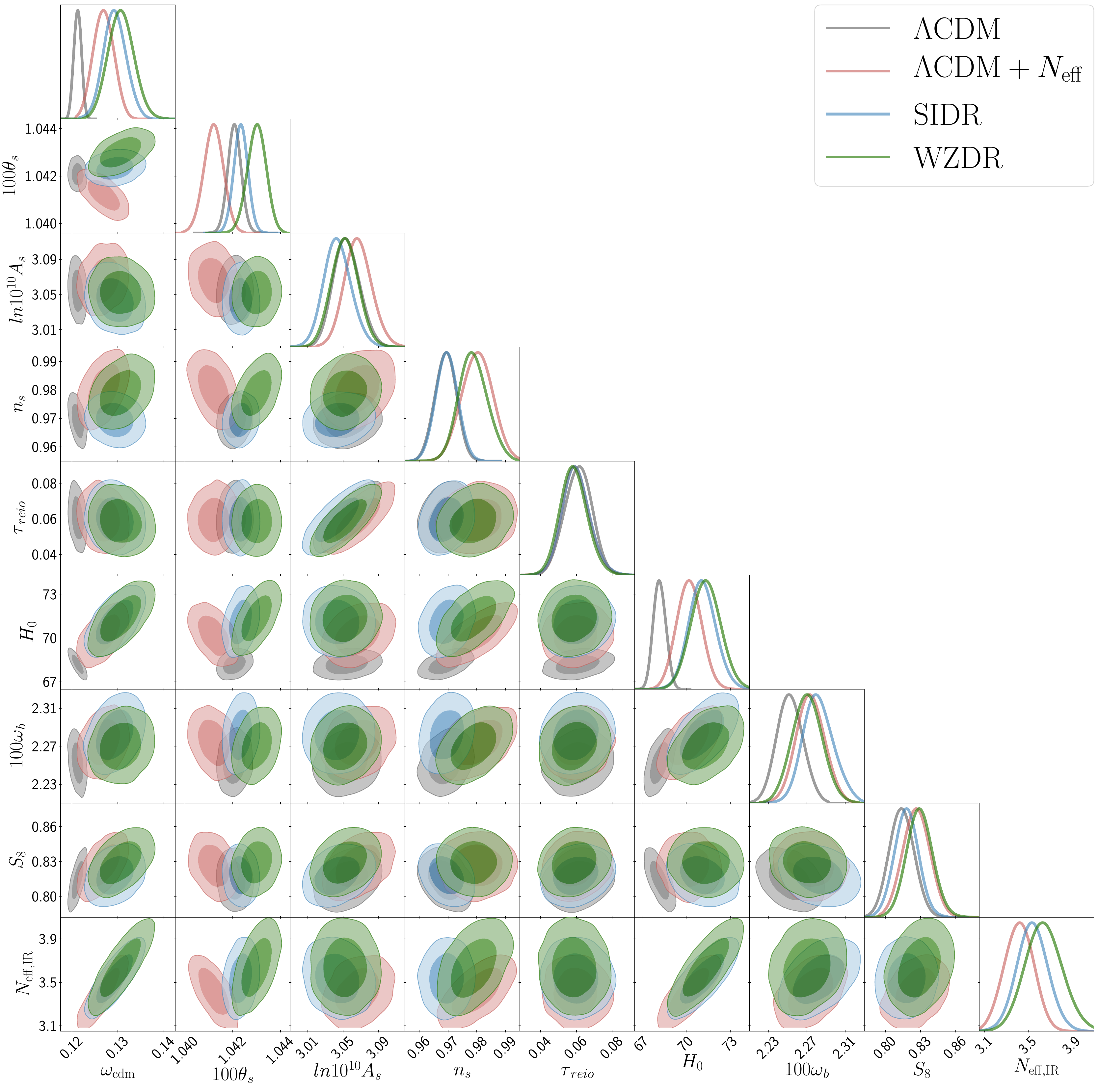}
	\caption{A comparison of the posteriors for models of a minimal stepped fluid (WZDR in green), conventional self-interacting dark radiation (SIDR in blue), additional free-streaming radiation ($\Lambda \text{CDM} + \Neff$ in red), and \lcdm\ (gray) when fitting to the dataset $\D+$ that includes SH0ES. The dark and light shaded regions correspond to $68.3\%$ and $95.4\%$ C.L., respectively. Most notable is the fact that WZDR predicts $H_0$ and $N_\text{eff,IR}$ posteriors that extend out to larger values, reflecting its ability to more successfully resolve the Hubble tension.}
	\label{fig:4model-triangle}
\end{figure*}

In this Appendix, we provide triangle plots and tables (shown below), which show the full set of posteriors and parameter values as determined from our analysis, respectively. In \Fig{wzdr-4param-triangle}, we compare the posteriors for the WZDR and General StepDR dark fluid models when fitting to the dataset $\D+$ including SH0ES. In \Fig{wzdr-SH0ES-noSH0ES-triangle}, we compare the posteriors for the WZDR model when fitting to a dataset that does ($\D+$) or does not ($\D$) include SH0ES. In \Fig{4model-triangle}, we compare the posteriors for the WZDR, General StepDR, SIDR, $\Lambda \text{CDM} + \Neff$, and \lcdm\ models when fitting to the dataset $\D+$. \Tab{fitresults} shows $\chi^2$ values as well as other fit parameters and measures relevant to evaluating the success of various models in resolving the Hubble tension. Furthermore, Tables~\ref{tb:meanD} and \ref{tb:meanDp} provide  posterior mean-value and $\pm 1 \sigma$ ranges for various cosmological parameters when fitting to either $\D$ or $\D+$, whereas Tables~\ref{tb:bestfitD} and \ref{tb:bestfitDp} show best-fit values.

\FloatBarrier

\begin{table*}[h!]
	\centering
	\hspace*{-1.5cm}
	\begin{tabular}{|c || c | c | c | c|c|c | c|c| c | c | c | c | c |} 
		\hline
		Model& $\Delta N_{param}$& $M_b + 1 \sigma$ & GT & $\QDMAP$
		& min $\chi^2_{\D+}$& min $\chi^2_{\D}$& $\Delta \chi^2_{{\cal D+}}$ &  $\Delta \text{AIC}$& &  $H_0^{BF}$ $\D$
		&  $H_0^{Mean}$ $\D$&  $H_0^{BF}$ ${\cal D+}$&  $H_0^{Mean}$ ${\cal D+}$ \\
		\hline\hline
		\lcdm & 0& $-19.418 + 0.011$ & 4.5& 4.5& 3828.3& 3808.5& 0.00& 0.00& & 67.6& 67.6& 68.2& 68.2\\
		$\Lambda \text{CDM} + \Neff$& 1& $-19.396 + 0.017$ & 3.7& 3.8& 3822.6& 3808.5& $-5.7$ & $-3.7$ & & 67.7& 68.4& 70.0& 70.2\\
		SIDR (ET) & 1& $-19.381 + 0.021$ & 3.2& 3.3& 3819.1& 3808.4& $-9.2$ & $-7.2$ & & 67.8& 68.9& 70.6& 70.8\\
		SIDR & 1& $-19.380 + 0.023$ & 3.1& 3.1& 3817.7& 3808.4& $-10.6$ & $-8.6$ & & 67.9& 68.9& 71.0& 71.0\\
		WZDR (ET) & 2& $-19.376+ 0.024$ & 3.0& 2.8& 3815.3& 3807.6& $-13.0$ & $-9.0$ & & 68.7& 69.0& 71.0& 70.9 \\
		WZDR& 2& $-19.368 + 0.027$ & 2.7& 2.4& 3813.2& 3807.3& $-15.1$ & $-11.1$ & & 69.1& 69.3& 71.4& 71.4\\
		General StepDR & 4& $-19.382 + 0.041$ & 2.5& 2.4& 3811.7& 3805.9& $-16.6$ & $-8.6$ & & 67.9& 68.8& 71.5& 71.5\\
		\hline
	\end{tabular}
	\hspace*{-1.5cm}
	\caption{A summary of the fit results for various models. The two versions of SIDR and WZDR refer to different assumptions about when the extra radiation is populated. ``SIDR (ET - Early Thermalization)" and ``WZDR (ET - Early Thermalization)" refer to models in which the energy density of interacting radiation is present during BBN, whereas ``SIDR" and ``WZDR" refer to the corresponding model in which we assume that this radiation was not yet present.  In the last four columns, $H_0$ is given in units of $\ksm$.}
	\label{tb:fitresults}
\end{table*}    

\begin{table*}[h!]{ }
	\centering
	\begin{tabular}{|c | c | c | c | c | c |} 
		\hline &&&&&\\[-8pt]
		& $\Lambda{\rm CDM}$ & $ \Lambda{\rm CDM} + \Neff$ & SIDR &  WZDR & General StepDR  \\
		\hline &&&&&\\[-8pt]
		100$\theta_s$ & $1.04195 ^{+0.00028}_{-0.00027}$ & $1.04166^{+0.00039}_{-0.00034}$ & $1.04208^{+0.00032}_{-0.00031}$ & $1.04243^{+0.00040}_{-0.00044}$ & $1.04339^{+0.00080}_{-0.00095}$  \\  [3pt]
		$\Omega_b h^2$ & $0.02240^{+0.00013}_{-0.00014} $ & $0.02248^{+0.00015}_{-0.00016}$  & $0.02255^{+0.00016}_{-0.00018}$ &  $0.02253^{+0.00015}_{-0.00017}$ &$ 0.02249^{+0.00019}_{-0.00020} $   \\ [3pt]
		$ \Omega_{\rm cdm}h^2$ & $ 0.11931^{+0.00091}_{-0.00094}$ & $0.1213^{+0.0013}_{-0.0022}$ & $0.1226^{+0.0019}_{-0.0032} $ & $0.1239^{+0.0023}_{-0.0037}$ & $0.1227^{+0.0037}_{-0.0042} $\\ [3pt]
		$\ln 10^{10} A_s$  &	$3.049^{+0.013}_{-0.016} $ & $3.054^{+0.015}_{-0.016} $ & $3.046 \pm 0.015$ & $3.050^{+0.014}_{-0.016} $ &$3.038^{+0.021}_{-0.019}$ \\ [3pt]
		$n_s$ & $0.9662^{+0.0040}_{-0.0035}$& $0.9703^{+0.0045}_{-0.0055} $ & $0.9670 \pm 0.0038 $ & $0.9721^{+0.0048}_{-0.0054} $ &$0.9664^{+0.0083}_{-0.0092}$\\ [3pt]
		$\tau_{\rm reio}$ &$0.0571^{+0.0069}_{-0.0078}$ & $0.0570^{+0.0071}_{-0.0077} $ & $0.0578^{+0.0075}_{-0.0080} $ & $0.0577^{+0.0068}_{-0.0080} $ &$0.0573^{+0.0069}_{-0.0078}$ \\ [3pt]
		$\nfs  $  & - & $3.17 ^{+0.03}_{-0.12} $ & $3.044 $& $3.044 $& $2.75^{+0.25}_{-0.25} $  \\ [3pt]
		$\nuv $  &-&-&-& $0.22^{+0.09}_{-0.17} $ & $0.39^{+0.15}_{-0.30} $  \\ [3pt]			
		$\nir $  &- &-& $0.19^{+0.05}_{-0.19} $ & $0.28^{+0.12}_{-0.22} $ & $0.48^{+0.18}_{-0.30} $  \\ [3pt]
		$\log_{10} (\zt)$  &- &- &- & $4.29^{+0.17}_{-0.16} $ & $4.35^{+0.18}_{-0.11} $ \\ [3pt] \hline  &&&&&\\[-8pt]	
		$M_b $ & ${-19.418}^{+0.011}_{-0.012} $ & ${-19.396}^{+0.017}_{-0.025}$ & ${-19.380}^{+0.023}_{-0.036}$ & ${-19.368}^{+0.026}_{-0.038}$ & ${-19.382}^{+0.041}_{-0.045} $ \\ [3pt]
		$H_0$ [km/Mpc/s] & $67.6 \pm 0.4 $ &  $68.4^{+0.6}_{-0.8} $ & $68.9^{+0.8}_{-1.2} $ & $69.3^{+0.9}_{-1.3}$ & $68.8^{+1.3}_{-1.5} $ \\ [3pt]
		$S_8$ & $0.825  \pm 0.010 $& $0.828 \pm 0.011 $ & $0.824 \pm 0.011 $ & $0.829 \pm  0.011 $ &$0.823 \pm 0.013$  \\  [3pt] \hline
		
	\end{tabular}
	\caption{Mean and $\pm 1 \sigma$ values for a fit to dataset $\D$.}
	\label{tb:meanD}
\end{table*}

\begin{table*}[h!]{ }
	\centering
	\begin{tabular}{|c | c | c | c | c | c |} 
		\hline &&&&&\\[-8pt]
		& $\Lambda{\rm CDM}$ & $ \Lambda{\rm CDM} + \Neff$ & SIDR &  WZDR & General StepDR  \\
		\hline &&&&&\\[-8pt]
		100$\theta_s$ & $1.04207 ^{+0.00029}_{-0.00031}$ & $1.04120 \pm 0.00042$ & $1.04233^{+0.00029}_{-0.00030}$ & $1.04300^{+0.00039}_{-0.00040}$ & $1.04324^{+0.00088}_{-0.00096}$  \\  [3pt]
		$\Omega_b h^2$ & $0.02252\pm 0.00014 $ & $0.02274^{+0.00016}_{-0.00015}$  & $0.02282 \pm  0.00016 $ &  $0.02270^{+0.00015}_{-0.00016}$ &$ 2.274^{+0.018}_{-0.017} $   \\ [3pt]
		$ \Omega_{\rm cdm}h^2$ & $ 0.11817^{+0.00097}_{-0.00092}$ & $0.1245^{+0.0026}_{-0.0025}$ & $0.1269^{+0.0029}_{0.0028} $ & $0.1288^{+0.0030}_{-0.0033}$ & $0.1294^{+0.0032}_{-0.0033} $\\ [3pt]
		$\ln 10^{10} A_s$  &	$3.054^{+0.015}_{-0.016} $ & $3.067^{+0.016}_{-0.017} $ & $3.044^{+0.015}_{-0.017} $ & $3.052 \pm 0.015 $ &$3.045 \pm 0.020 $ \\ [3pt]
		$n_s$ & $0.9691^{+0.0037}_{-0.0039}$& $0.9802^{+0.0057}_{-0.0055} $ & $0.9694^{+0.0037}_{-0.0036} $ & $0.9789^{+0.0047}_{-0.0052} $ &$0.9752^{+0.0082}_{-0.0086}$\\ [3pt]
		$\tau_{\rm reio}$ &$0.0606^{+0.0071}_{-0.0086}$ & $0.0598^{+0.0075}_{-0.0082} $ & $0.0599^{+0.0070}_{-0.0084} $ & $0.0587^{+0.0075}_{-0.0077} $ &$0.0585^{+0.0071}_{-0.0078}$ \\ [3pt]
		$\nfs   $  &- & $3.41 ^{+0.13}_{-0.15} $ & $3.044$ & $3.044 $& $2.93 \pm 0.27 $  \\ [3pt]
		$\nuv $  &-&-&-& $0.46^{+0.13}_{-0.14} $ & $0.601^{+0.24}_{-0.27} $  \\ [3pt]			
		$\nir $  &- &-& $0.47^{+0.15}_{-0.14} $ & $0.60^{+0.16}_{-0.18}$ & $0.74^{+0.26}_{-0.29} $  \\ [3pt]
		$\log_{10} (\zt)$  &- &- &- & $4.26^{+0.12}_{-0.13} $ & $4.30^{+0.09}_{-0.13} $ \\ [3pt] \hline	 &&&&&\\[-8pt]
		$M_b $ & ${-19.404} \pm 0.012 $ & $ {-19.342} \pm 0.026 $ & $ {-19.320}^{+0.027}_{-0.028}$ & $ {-19.308} \pm 0.029$ & ${-19.302}^{+0.031}_{-0.029} $ \\ [3pt]
		$H_0$ [km/Mpc/s] & $68.2 \pm 0.4 $ &  $70.2 \pm 0.9 $ & $71.0^{+0.9}_{-1.0} $ & $71.4 \pm 1.0 $ & $71.5^{+1.1}_{-1.0} $ \\ [3pt]
		$S_8$ & $0.814  \pm 0.011 $& $0.827 \pm 0.012 $ & $0.818 \pm 0.010 $ & $0.829 \pm 0.011 $ &$0.823 \pm 0.013$  \\  [3pt] \hline
		
	\end{tabular}
	\caption{Mean and $\pm 1 \sigma$ values for a fit to dataset $\D+$.}
	\label{tb:meanDp}
\end{table*}

\begin{table*}[h!]{ }
	\centering
	\begin{tabular}{|c | c | c | c | c | c |} 
		\hline &&&&&\\[-8pt]
		& $\Lambda{\rm CDM}$ & $ \Lambda{\rm CDM} + \Neff$ & SIDR &  WZDR & General StepDR  \\
		\hline &&&&&\\[-8pt]
		100$\theta_s$ &1.04192&1.04193&1.04196&1.04236&1.04299  \\  [3pt]
		$\Omega_b h^2$ &0.022406&0.022410&0.022450&0.022533& 0.022416   \\ [3pt]
		$ \Omega_{\rm cdm}h^2$ & 0.11930&0.11930&0.11980&0.12320&0.12037 \\ [3pt]
		$\ln 10^{10} A_s$  &3.049 &3.049&3.049&3.049& 3.045 \\ [3pt]
		$n_s$ & 0.9666&0.9666&0.9674&0.9724&0.9693 \\ [3pt]
		$\tau_{\rm reio}$ &0.0571 &0.0571&0.0572&0.0565&0.0573 \\ [3pt]
		$\nfs   $  &3.044 & 3.044 &3.044&3.044&2.85 \\ [3pt]
		$\nuv $  &-&-&-& 0.18 & 0.16 \\ [3pt]			
		$\nir $  & -&-& 0.03 & 0.23 & 0.23   \\ [3pt]
		$\log_{10} (\zt)$  &- &- &- &4.30 &4.46  \\ [3pt] \hline  &&&&&\\[-8pt]	
		$M_b $ &${-19.418}$ &${-19.417}$ &${-19.411}$&${-19.375}$& ${-19.409}$\\ [3pt]
		$H_0$ [km/Mpc/s] & 67.6 &67.7 &67.9&69.1& 67.9\\ [3pt]
		$S_8$ &0.825&0.825&0.825 &0.828& 0.827 \\  [3pt] \hline &&&&&\\[-8pt]
		$\chi^2_{\rm tot}$  & 3808.5 & 3808.5 & 3808.4 &  3807.3 & 3805.9\\ 
		\hline
		
	\end{tabular}
	\caption{Best-fit values for a fit to dataset $\D$.}
	\label{tb:bestfitD}
\end{table*}

\begin{table*}[h!]{ }
	\centering
	\begin{tabular}{|c | c | c | c | c | c |} 
		\hline &&&&&\\[-8pt]
		& $\Lambda{\rm CDM}$ & $ \Lambda{\rm CDM} + \Neff$ & SIDR &  WZDR & General StepDR  \\
		\hline &&&&&\\[-8pt]
		100$\theta_s$ & $1.04204 $& $1.04129$ & $1.04234$ &$ 1.04303$ &$1.04331$  \\  [3pt]
		$\Omega_b h^2$ &$ 0.022531$& $0.022712$ & $0.022817$ & $0.022718$ &$0.022726 $ \\ [3pt]
		$ \Omega_{\rm cdm}h^2$ &$ 0.1183$0 & $0.12372 $&$ 0.12681$ & $0.12880 $&$0.12952$\\ [3pt]
		$\ln 10^{10} A_s$  &$	3.052$ & $3.066$ &$ 3.043$ & $3.053$ &$3.045$ \\ [3pt]
		$n_s$ & $0.9692$&$ 0.9798$ & $0.9693 $&$0.9801$ &$0.9777$ \\ [3pt]
		$\tau_{\rm reio}$ &$0.0597$& $0.0599 $&$ 0.0594 $& $0.0587$ &$0.0575$ \\ [3pt]
		$\nfs   $  &$3.044 $&$ 3.37$ &$ 3.044$ & $3.044$ & $2.94 $ \\ [3pt]
		$\nuv $  &-&-&-&$0.46$ &$0.57 $ \\ [3pt]			
		$\nir $  &- &-& $0.47$ &$0.59$&$0.72$  \\ [3pt]
		$\log_{10} (\zt)$  &- &- & -&$4.27$ &$4.33$\\ [3pt] \hline	 &&&&&\\[-8pt]
		$M_b $ &$ {-19.404}$ &$ {-19.348}$ &$  {-19.320}$ & $ {-19.307}$& ${-19.301}$\\ [3pt]
		$H_0$ [km/Mpc/s] & $68.2$& $ 70.0 $& 71.0& 71.4& 71.5\\ [3pt]
		$S_8$&$0.815$&$0.825$&$0.817$&$0.830$&$0.825$ \\  [3pt] \hline &&&&&\\[-8pt]
		$\chi^2_{\D}$  &$3809.9$ & $3814.5$&  $3813.5$ & $ 3810.3$ & $3809.37 $\\ [3pt]
		$\chi^2_{\rm tot}$  &$3828.3$&$ 3822.6$&  $3817.7$&  $3813.2 $& $3811.7 $\\ \hline
		
	\end{tabular}
	\caption{Best-fit values for a fit to dataset $\D+$. The value of $\chi^2_{\D}$ (in the second to last row) corresponds to the best-fit points from a fit to dataset $\D+$ but only includes  $\chi^2$ contributions from dataset $\D$. }
	\label{tb:bestfitDp}
\end{table*}

\end{document}